# Software Dependability Modeling Using An Industry-Standard Architecture Description Language

Ana-Elena Rugina[1,*], Peter H. Feiler[2], Karama Kanoun[1] and Mohamed Kaâniche[1]

1: LAAS–CNRS, University of Toulouse, Toulouse (France)
2: Software Engineering Institute, Carnegie Mellon University, Pittsburgh, PA (U.S.A.)

*: Contact author, now with EADS ASTRIUM, Ana-Elena.Rugina@astrium.eads.net,
31 Av. des Cosmonautes, 31402 Toulouse cedex 4, France

**Abstract:** Performing dependability evaluation along with other analyses at architectural level allows both making architectural tradeoffs and predicting the effects of architectural decisions on the dependability of an application. This paper gives guidelines for building architectural dependability models for software systems using the AADL (Architecture Analysis and Design Language). It presents reusable modeling patterns for fault-tolerant applications and shows how the presented patterns can be used in the context of a subsystem of a real-life application.

**Keywords:** AADL, fault tolerance, reuse, patterns

## 1. Introduction

Modeling software architectures has proved to be useful for promoting reuse and evolution of large applications using extensively components-off-the-shelf (COTS). In addition, performing several analyses of quality attributes such as dependability and performance on a common architectural model is particularly interesting, as this allows making architectural tradeoffs [1].

The AADL (Architecture Analysis and Design Language) [2] is a textual and graphical language that provides precise execution semantics for modeling the architecture of software systems and their target platform. It has received an increasing interest from the embedded safety-critical industry (e.g., Honeywell, Rockwell Collins, Lockheed Martin, the European Space Agency, Astrium, Airbus) during the last years. The AADL is characterized by all the properties that an architecture description language (ADL) should provide (composition, abstraction, reusability, configuration, heterogeneity, analysis) [3].

In this paper, we focus on architecture-based dependability modeling and evaluation using the AADL. Our work aims at helping engineers using the AADL for other purposes (e.g., for performance analyses), to integrate dependability modeling in their development process.

We provide guidance on using the AADL language for modeling behaviors of fault-tolerant software systems, and show that the development of patterns is very useful to facilitate the modeling of fault tolerance behavior and to enhance the reusability of the models. We define a *fault tolerance pattern* as a reusable model describing a fault tolerance strategy at the architectural level. To be used in a particular system, a pattern must be instantiated and customized if necessary.

The use of patterns and, more generally, dependability modeling at architectural level favors the reduction of recurrent dependability modeling work and the understandability of the dependability model (thus reflecting the modularity of the architecture) [4] and allows the designer to reason about fault tolerance and to assign exceptional behavior responsibilities among components [5]. At the same time, dependability measures (i.e., availability, reliability, safety) can be evaluated based on the AADL model. This allows predicting the effects of particular architectural decisions on the dependability of the system [6]. Other analyses (e.g., related to performance) may be performed on the same AADL model, which allows understanding the tradeoff between the benefits of a certain fault tolerance pattern and its impact on the application's performance [7].

From a practical point of view, the AADL model must be transformed into a stochastic model such as a Markov chain [8] or a Generalized Stochastic Petri net [9], to obtain dependability measures such as reliability, availability, etc. In this paper we focus on the use of patterns to facilitate the AADL model construction.

The paper is organized as follows. Section 2 surveys related work. Section 3 outlines the main concepts of the AADL and its support for dependability modeling. Section 4 gives guidance, resulting from our experience, on building dependability models for fault-tolerant software systems using the AADL. Section 5 presents AADL fault tolerance patterns for three duplex software systems (i.e., dual-redundant systems), differing by their error detection mechanisms. Section 6 illustrates the use of patterns to model a real-life application and shows examples of dependability analysis results of interest for software engineers. Finally, conclusions and perspectives are presented in Section 7.

## 2. Related work

Software architecture modeling for dependability analysis and evaluation has received a growing interest during the last two decades. Early approaches have focused on the development of analytical models to analyze the sensitivity of the application reliability to the software structure and the reliabilities of its components (see e.g., [10, 11] and the survey presented in [12]). More recently, the emergence of component-based software engineering approaches and architecture description languages (ADLs) led to the proliferation of research activities on software architectures and methodologies allowing the analysis and evaluation of performance- and dependability-related characteristics. Significant efforts have been focused on the Unified Modeling Language (UML)[1]. In particular, a number of recent papers consider the transformation of UML specifications (enriched e.g., with timing constraints and dependability related information) into different types of analytical models (e.g., Petri nets [13, 14], dynamic fault trees [15]) used to obtain dependability or performance measures.

Besides UML, various ADLs have recently received increasing attention from industry and academia. A classification of software architecture description languages including a critical analysis of their modeling capabilities (in particular compared to UML) is presented in [16]. Among ADLs, the AADL/MetaH[2] provides advanced support for analyzing quality attributes. It also has substantial support for modeling reconfigurable architectures. These characteristics led to its serious consideration in the embedded safety-critical industry [17].

The AADL allows describing separately the analysis-related information that may be plugged into the architectural model. This feature enhances the reusability and the readability of the AADL architectural model that can be used as is for several analyses (formal verification [18], scheduling and memory requirements [19], resource allocation with the Open Source AADL Tool Environment (OSATE)[3], research of deadlocks and un-initialized variables with the Ocarina toolset[4]).

The reusability of the AADL model is also enhanced by the use of a set of fault tolerance patterns. The hot standby redundancy pattern presented in [7] has been a source of inspiration for the three patterns presented in this paper. The pattern presented in [7] aims at easing the understanding of the functional architecture by clearly showing what is replicated and what the active system components are. Our patterns additionally include a customizable layer of dependability-related information (error/failure and recovery behavior) and of dynamics necessary for evaluating dependability measures. The proposed patterns can be used in the context of the iterative dependency-driven modeling approach presented in [20].

Our work complements other existing initiatives that investigated the development of fault tolerance patterns based on object-oriented approaches and UML ([21-23]) or other languages ([24]).

---

[1] http://www.uml.org

[2] MetaH is a prototype of the AADL developed by Honeywell under US Government sponsorship (DARPA and others) to prove the concept.

[3] http://www.aadl.info/OpenSourceAADLToolEnvironment.html

[4] http://ocarina.enst.fr

## 3. Overview of the AADL language

In the AADL, systems are composite components modeled as hierarchical collections of interacting application components (processes, threads, subprograms, data) and a set of compute platform components (processors, memory, buses, devices). The application components are bound to the compute platform. Dynamic aspects of system architectures are captured with the AADL operational mode concept. Different operational modes of a system or system component represent different system configurations and connection topologies, as well as different sets of property values.

Each AADL system component has two levels of description: the *component type* and the *component implementation*. The component type describes how the environment sees that component (i.e., its properties and features). Examples of features are *in* and *out* ports that represent directional access points to the component. The AADL defines three types of ports: event, data, and event data (modeling respectively flows of control, data, and control and data). One or more component implementations can be associated with the same component type, matching different component implementation structures in terms of subcomponents, connections and operational modes.

An AADL architectural model can be annotated with dependability-related information (such as faults, failure modes and repair actions, error propagation, etc.) through the standardized Error Model Annex [25]. *AADL error models* allow modeling complex and realistic components' behaviors in the presence of faults, as shown in [26]. Generic error models are defined in libraries and are associated with application components, compute platform components, as well as the connections between them. When a generic error model is associated with a component, it can be customized if necessary by setting component-specific values for the arrival rate or the probability of occurrence for error events and error propagations declared in the error model.

Error models consist of two parts: the *error model type* and the *error model implementation*. The error model type declares a set of `error states`, `error events` (inherent to the component) and `error propagations`[5]. These items can be customized when the error model is associated with a specific component. `Occurrence` properties specify the arrival rate or the occurrence probability of events and propagations. The error model implementation declares `error transitions` between states, triggered by events and propagations declared in the error model type.

Figure 1 shows an error model of a component that may fail and that is restarted to regain its error free state. The component cannot be influenced by propagations coming from its environment, as it does not declare `in` propagations. An `out` propagation is used to indicate notification of dependent components when the component fails.

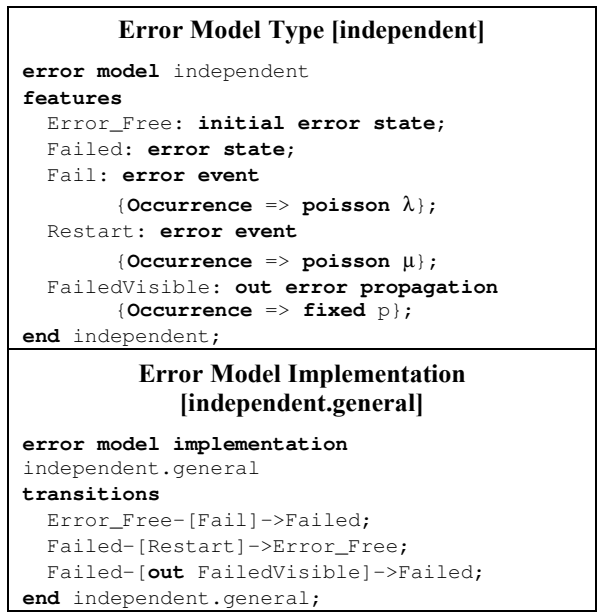

**Figure 1. Two-state error model**

Interactions between the error models of different components are determined by interactions between components of the architectural model through connections and bindings. For example, if a component has an outgoing port connection to another component, then its `out` propagation for that port gets mapped to the name-matching `in` propagation declared in the error model of the receiving component. In some cases, it is desirable to model how error propagations from multiple sources are handled. This is modeled by specifying filters and masking conditions for propagations, using `Guard` properties associated with features. The interested reader can refer to [26] for an extensive list of generic reusable error models and `Guard` properties.

---

[5] Note that `error states` can model error-free states, `error events` can model repair events and `error propagations` can model all kinds of notifications. In this paper we refer to `error states`, `error events`, `error propagations` and `error transitions` without the qualifying term `error` in contexts where the meaning is unambiguous.

### 4. Guidelines for modeling dependability

In order to analyze the dependability of an application at architectural level, it is necessary to enrich the architectural model with dependability-related information relevant to the targeted measure(s). Generally, dependability models include fault assumptions, stochastic parameters for the system, description of recovery and fault tolerance mechanisms, and characteristics of phases in a phased-mission system. An AADL user describes a system's architecture in the AADL and annotates this architectural model with error models containing relevant dependability-related information.

Section 4.1 discusses the role of AADL operational modes and mode transitions. Section 4.2 discusses the use of operational modes versus error states. Section 4.3 presents mechanisms for representing the logic connecting error states to operational modes.

### 4.1. What operational modes are good for

An *operational mode*[6] is an operational state of an AADL component. Exactly one mode is the initial mode. A component is in one mode at a time. Mode transitions model dynamic operational behavior, i.e., switching between configurations of subcomponents and connections. Mode-specific property values characterize quality attributes for each operational mode. A mode transition may be triggered by: (1) an `out` (or `in out`) event port of a subcomponent of the component declaring the modes; (2) an `in` (or `in out`) event port of the component itself; (3) a local event port[7].

A mode transition is triggered by any event that arrives through the port named in the mode transition. A mode transition may list multiple event ports as its trigger condition. In such a case, an event through any of the ports triggers the transition (an `or` logic is assumed). For dependability analyses, more advanced event-based mode transition conditions (reflecting e.g. voting protocols) can be specified through `Guard_Event` and `Guard_Transition` properties, as presented in Section 4.3. The state machine formed by modes and mode transitions in a system or system component implementation must be deterministic.

---

[6] We will further refer to operational modes simply as modes.
[7] The errata document to the AADL standard provides the ability to declare such local ports, representing a call to a pre-declared Raise_Event subprogram in a thread.

4.2. Operational modes vs. error states

Modes of operation in phased-mission systems, as well as fault-tolerant configurations are modeled by AADL modes.

*Operational modes in phased-mission systems* model configurations representative of different phases in a mission. For example, in the case of an aircraft model, one may distinguish between the takeoff, cruise and landing phases. During each phases, the system would have a particular configuration with active components and connections. Also, different types of faults may affect the system in different phases.

*Mode-specific fault-tolerant system configurations reflect* the fault tolerance strategy chosen for the system or for particular parts of the system. For example, a fault-tolerant system formed of three components may have a nominal operational mode corresponding to a triple-redundancy configuration and another degraded operational mode corresponding to a duplex configuration.

Usually, phased-mission systems also need modes to represent fault tolerance mechanisms. In the AADL, this nesting of modes is captured by phased-mission modes in a component, which is a subcomponent of a system component whose modes represent alternative configuration of its redundant subcomponents.

The difference between modes and error states lies primarily in their semantics. Error states result from occurrences of error events (faults, repair events) while modes represent operational states of the system that may be totally independent of the occurrence of error events.

4.3. Connecting error states to modes

The AADL allows us to model logical error states separate of the operational mode of the running application. It also allows connecting logical error states and operational modes by translating logical error states into actions (under the form of architectural events) on the running system through `Guard_Event` properties. `Guard_Event` properties map error state configurations into architectural events that are sent through ports and thus may affect the behavior of receiving components by triggering mode transitions. An architectural event arriving through a port will unconditionally trigger a mode transition.

Sometimes one may need to constrain a mode transition in a system component to reflect specific conditions such as a voting protocol to decide on fault handling. This can be achieved through the use of `Guard_Transition` properties associated with mode transitions and specifying mode transition logic expressions overriding the default `or` condition on events arriving through ports named in the mode transition.

`Guard_Event` and `Guard_Transition` properties can be used as advanced decision mechanisms that drive reconfiguration strategies.

## 5. Fault tolerance patterns

In this section, we consider a fault-tolerant duplex system that uses the hot standby redundancy scheme. We present successively, in sections 5.1, 5.2 and 5.3, patterns for three architectural implementations of this system. Section 5.4 gives concluding remarks.

Each of the three models contains two identical active components, *Comp1* and *Comp2*, modeled as threads. Both threads process the event and data input stream received by the redundant system through the port *sysInput* but only one component's output is made visible as output of the redundant system through the `out` event data port *sysOutput*. The three patterns differ in terms of their error detection mechanisms, as follows.

The first pattern models the error detection by intermediate checkpoints between the two components. The direction of the data flow is from the primary component to the one that is in standby. The redundant system has two operational modes, one in which *Comp1* is the primary and another one in which *Comp2* is the primary. The component in standby monitors the checkpoints sent by the primary and decides to take over and change the operational mode of the system, if it detects a failure.

In the second pattern, the error detection is achieved by a separate monitoring and control component, *Controller*. The outputs of both active components are connected to the output of the system, *sysOutput*, but only one component provides the output at a given instant in time. This is modeled by two modes associated with each active component. Only the component in mode *primary* sends data to the output of the system. The *Controller* initiates the mode transitions.

The third pattern models error detection by mutual observation of outputs. The outputs of both active components are connected to the output of the system. Each of the two active components monitors the output of its sibling and decides whether to provide the output or not.

In the first pattern (detection by intermediate checkpoints), modes are represented at the system's level. In the two other patterns, modes are represented at the component level.

Each of the three patterns is formed of an AADL architectural model and of error model annex subclauses that associate the dependability-related information to the components of the architectural model. These subclauses may be further refined during the development cycle to detail the internal behavior of components and component-specific occurrence of propagations.

5.1. Detection by intermediate checkpoints

Figure 2 shows the AADL architectural model for this system implementation, using the AADL graphical notation.

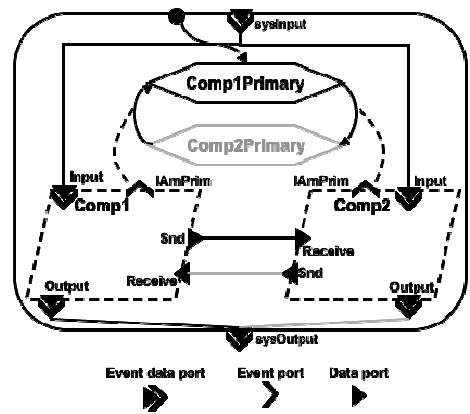

Figure 2: Detection by checkpoints (graphical)

The system has two operational modes. In mode *Comp1Primary*, *Comp1*'s output is made visible as output of the system. In mode *Comp2Primary*, *Comp2*'s output is made visible as output of the system. Thus, the connection from *Comp1* to the `out` event data port *sysOutput* of the system is active in mode *Comp1Primary* while the connection from *Comp2* to the `out` event data port *sysOutput* of the system is active in mode *Comp2Primary*.

Based on the input from the other component and on its own state, a component in standby can decide to take over by initiating a mode transition. Thus, the transition from *Comp1Primary* to *Comp2Primary* is triggered by the `out` event port *IAmPrim* of *Comp2*. If both components fail successively and if their failures are detectable, the first one restarted becomes the *primary*.

For this pattern, we show the complete AADL architectural model of the system and the error model annex subclause associated with its components in Figure 3.

```
thread software
features
  Snd: out data port;
  Receive: in data port;
  Input: in event data port;
  Output: out event data port;
  IAmPrim: out event port;
end software;

thread implementation software.generic
annex Error_Model {**
      Model => independent.general;

      Guard_Event =>
(Receive[FailedVisible] and self[Error_Free])
                  applies to IAmPrim;
**};
end software.generic;
```

```
system HotStandBy
features
  sysInput: in event data port;
  sysOutput: out event data port;
end HotStandBy;

system implementation HotStandBy.checkpoints
subcomponents
  Comp1: thread software.generic;
  Comp2: thread software.generic;
connections
  data port Comp1.Snd->Comp2.Receive
      in modes Comp1Primary;
  data port Comp2.Snd->Comp1.Receive
      in modes Comp2Primary;
  event data port sysInput->Comp1.Input;
  event data port sysInput->Comp2.Input;
  event data port Comp1.Output->sysOutput
      in modes Comp1Primary;
  event data port Comp2.Output->sysOutput
      in modes Comp2Primary;
modes
  Comp1Primary: initial mode;
  Comp2Primary: mode;
  Comp1Primary -[Comp2.IAmPrim]->Comp2Primary;
  Comp2Primary -[Comp1.IAmPrim]->Comp1Primary;
end HotStandBy.checkpoints;
```

Figure 3. Detection by checkpoints (textual)

The threads have the same component type and implementation, given in the upper part of Figure 3. The lower part shows the component type and implementation of the system.

`Guard_Event` properties specify the mode switching conditions. The system must switch from *Comp1Primary* to *Comp2Primary* when *Comp2* detects the failure of *Comp1* or when both components failed and *Comp2* was restarted first (i.e., *Comp2* is *Error_Free* and *Comp1* sends the `out` propagation *FailedVisible*).

5.2. Detection by a separate controller

The system modeled in Figure 4 consists of two identical active threads (*Comp1* and *Comp2*) and one controller component (*Controller*).

Each of the active components can be in one of two modes: *primary* and *standby*. When a component is in *primary* mode, it provides the service expected from the redundant system. The *Controller* monitors the two components. If it detects the failure of the one in *primary* mode, it initiates a mode switch in each component, so that the one that is *Error_Free* continues to provide the service. Also, if it detects the failure of both components, it waits until one of them becomes operational and orders it to go to *primary* mode.

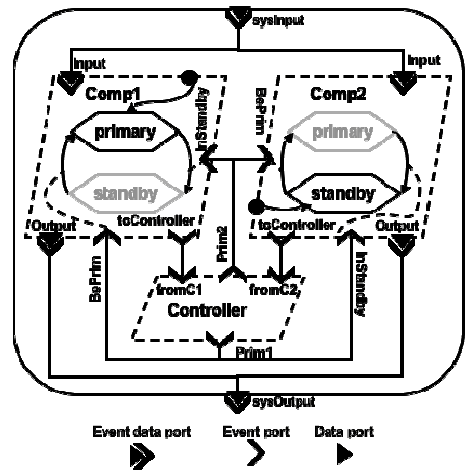

Figure 4: Separate controller (graphical)

Figure 5 shows the textual AADL architectural models and error model annex subclauses of the component that is initially in *primary* mode (upper part of the figure) and of the controller (lower part of the figure).

```
thread software
features
  InStandby: in event port;
  BePrim: in event port;
  Input: in event data port;
  Output: out event data port;
  toController: out event port;
end software;
---------------------------------------
thread implementation software.primary
modes
  primary: initial mode;
  standby: mode;
  primary-[InStandBy]->standby;
  standby-[BePrim]->primary;
annex Error_Model {**
      Model => independent.general;
      Occurrence => fixed 0.9
            applies to error FailedVisible;
      Occurrence => poisson 1e-3
            applies to error Fail;
**};
end software.primary;

system controller
features
  fromC1, fromC2: in event port;
  Prim1, Prim2: out event port;
end controller;
---------------------------------------
system implementation controller.generic
annex Error_Model {**
      Model => independent.general;
      Occurrence => poisson 1e-6
                applies to error Fail;

      Guard_Event =>
         fromC1[FailedVisible] and
         fromC2[Error_Free] and
         self[Error_Free]
                applies to Prim2;

      Guard_Event =>
         fromC2[FailedVisible] and
         fromC1[Error_Free] and
         self[Error_Free]
                applies to Prim1;
**};
end controller.generic;
```

Figure 5. Separate controller (textual)

5.3. Detection by mutual observation

Figure 6 shows the architectural model of this system. Each of the two active components can be in one of these three modes: *primary*, *standby* and *reboot*. Initially, one component is in *primary* mode while the other one is in *standby*. When a component is in *primary* mode, it provides the service expected from the redundant system. The two components observe each other's outputs. Based on these observations and on its own state, each component decides whether it must be the sender of output. When a failure occurs in a component, the component goes to *reboot* mode. If the failed component was in *primary* mode, the other component should take over so that the service expected continues to be provided. If both components fail one after the other, the first one restarted becomes the *primary*.

In the two previous patterns, the mode transitions of a component or system are controlled by its subcomponents or by a separate controller. In this pattern, we use self-managing components that control their own mode transitions. This is modeled by local event ports (represented as dotted ovals) triggering the mode transitions.

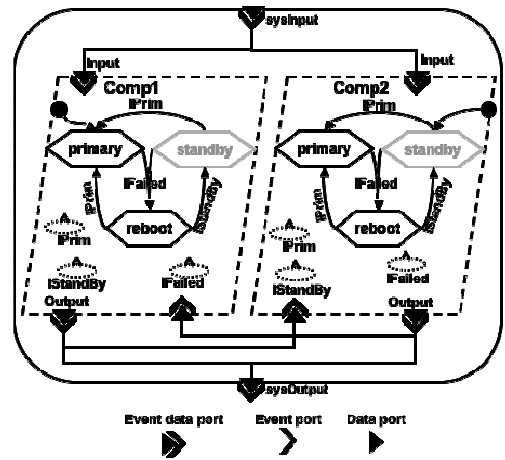

Figure 6. Mutual observation (graphical)

The two components' architectural models are identical except for their initial modes, i.e., one is initially in *primary* mode while the other one is in *standby* mode. Thus, Figure 7 shows only the textual AADL architectural model of the component that is initially in *primary* mode and its associated error model annex subclause.

In Figure 7, `Guard_Event` properties are associated with all internal ports (expressed by **self**.*eventname*) named in mode transitions. For example, the first declared `Guard_Event` property specifies that the component moves to *reboot* mode when it fails.

5.4. Concluding remarks

In sections 5.1, 5.2 and 5.3, we assume that all components (*Comp1*, *Comp2*, *Controller*) have the same behavior in the presence of faults (represented by the error model of Figure 1). More complex behavior can be considered for each component, by customizing the patterns.

This is achieved by changing the `Model` property in the error model annex subclause associated with a particular component.

Other patterns, modeling different fault tolerance schemes and impacting the dependability of the system, are presented in [26].

```
thread software
features
  Input: in event data port;
  Output: out event data port;
  fromReplica: in event data port;
end software;
─────────────────────────────────────
thread implementation software.primary
modes
  primary: initial mode;
  standby, reboot: mode;
  primary-[self.IFailed]->reboot;
  standby-[self.IFailed]->reboot;
  reboot-[self.IPrim]->primary;
  reboot-[self.IStandby]->standby;
  standby-[self.IPrim]->primary;
annex Error_Model {**
      Model => independent.general;
      Occurrence => fixed 0.9
                applies to error
FailedVisible;
      Guard_Event => self[Failed]
                applies to self.IFailed;
      Guard_Event =>
      fromReplica[FailedVisible]
           and self[Error_Free]
                applies to self.IPrim;
      Guard_Event =>
      fromReplica[Error_Free] and
           self[Error_Free]
                applies to self.IStandby;
**};
end software.primary;
```

Figure 7. Mutual observation (textual)

### 6. Application example

We illustrate the use of AADL architectural patterns for dependability analyses on a safety-critical subsystem of the French Air Traffic Control System. This system has been studied in [27] using generalized stochastic Petri nets (GSPN) for comparing candidate architecture solutions, with respect to availability. The contribution of this paper is to show how to model it using the AADL fault tolerance patterns.

The subsystem is formed of two fault-tolerant distributed software units that are in charge of processing respectively flight plans (FPunit) and radar data (RDunit). Two processors can host these units. The subsystem must be highly available.

We consider two candidate architectures for this subsystem, referred to as *Configuration1* and *Configuration2*. Both of them use two processors. The two components of each fault-tolerant subsystem are bound to separate processors. The difference between the two configurations lies only in the bindings of RDunit threads to processors. In *Configuration1*, the thread that initially delivers the service (*Comp1*) is bound to *Processor2* while *Comp2* is bound to *Processor1*. In *Configuration2*, the bindings are the other way round (i.e., *Comp1* is bound to *Processor1* while *Comp2* is bound to *Processor2*). The thread that delivers the service in the FPunit exchanges data with the RDunit. Connections between threads bound to separate processors are bound to a bus whose failure causes the failure of the RDunit.

Figure 8 presents both candidate architectures using the AADL graphical notation. For the sake of clarity, we show the thread binding configurations in Figure 8-a and the bus and the connection bindings to the bus in Figure 8-b. We assume that the error detection is achieved through intermediate checkpoints (pattern 5.1).

The modeling effort is limited. We need to instantiate the chosen pattern, to connect together the two instances and to bind the threads to processors. Besides these rather simple actions, we need to associate an error model annex subclause with the bus connecting the two processors. The error model annex subclauses may be further customized, to consider particular reconfiguration strategies.

The AADL models of the two candidate architectures presented in Figure 8 are transformed into GSPN that are not shown in this paper due to space limitations (see [9] for details about the transformation process). Figure 9 gives, as an example of result obtained from the GSPN processing, the unavailability of the two candidate architectures. The varying parameter $\lambda_c$ is the occurrence rate of a bus failure. $\lambda_c \leq 10^{-6}$/h corresponds to a redundant bus. For *Configuration1*, the impact of this parameter is important when $\lambda_c \geq 10^{-5}$/h. *Configuration2* is much less influenced by $\lambda_c$, as in *Nominal* mode, the communication between the two units does not go through the bus. From a practical point of view, if $\lambda_c \geq 10^{-5}$/h, *Configuration2* is recommended. Otherwise the two candidate architectures are equivalent, from the availability viewpoint.

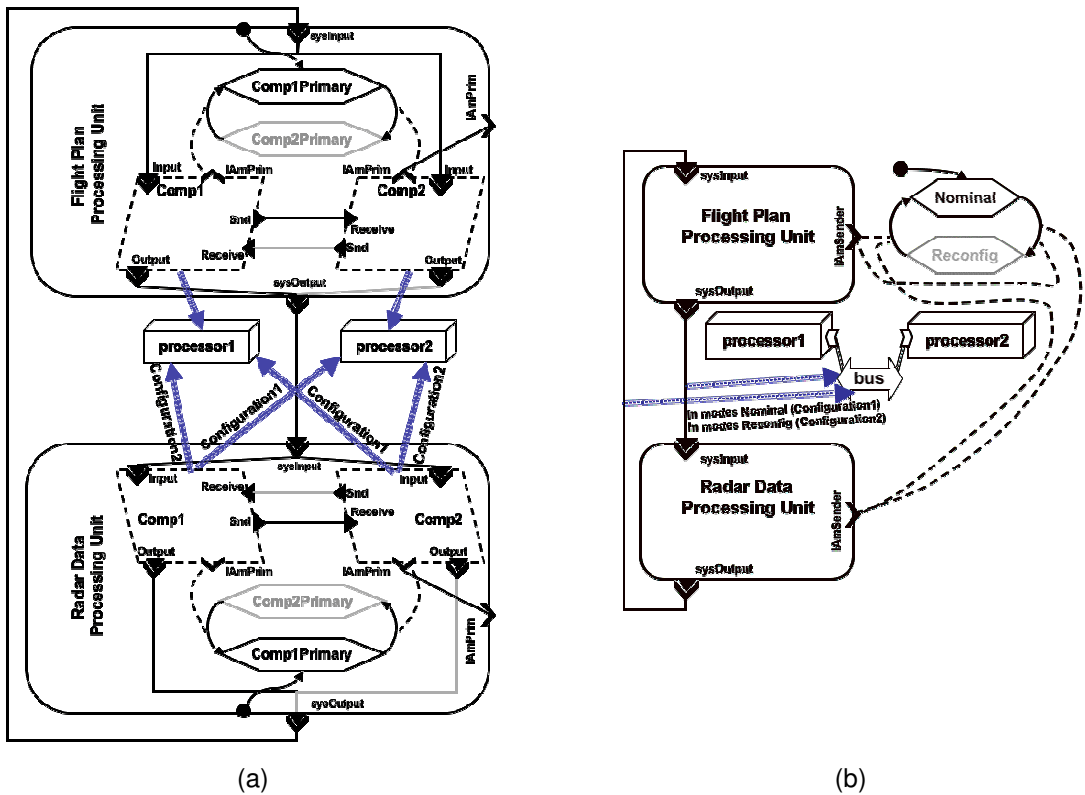

Figure 8: Models of Air Traffic Control System candidate architectures

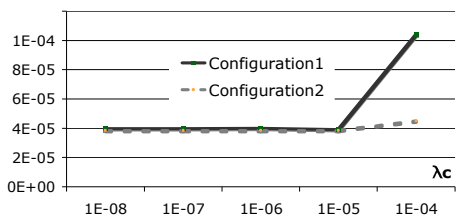

Figure 9: Unavailability

## 7. Conclusion

Performing several dependability and performance-related analyses on a same architectural model is particularly interesting for software engineers as having qualitative and quantitative information about a candidate architecture allows making architectural tradeoffs.

The AADL (Architecture Analysis and Design Language) is a mature industry-standard well suited to address quality attributes. This paper illustrated its use for dependability modeling of fault-tolerant systems. Model reusability is an essential issue in the context of complex safety-critical evolvable applications. We presented patterns for modeling fault-tolerant applications and we showed that they enhance the reusability and the understandability of the model. Finally, we showed a pattern-based example used for evaluating the availability of two candidate architectures for a subsystem of the French Air Traffic Control System.

Future extensions of this work include the construction of a library of AADL architectural patterns, and of error models to express common dependencies in dependable applications (i.e., restoration and reconfiguration strategies).

## 8. Acknowledgements

This work results from a cooperation between LAAS-CNRS and the Software Engineering Institute (SEI), initiated during the visit of Ana-Elena Rugina at the SEI. It has been partially financed by the ReSIST Network of Excellence (Resilience for Survivability in IST, IST-26764) and the European Social Fund.